\documentclass[twocolumn,english,pra, showpacs,superscriptaddress]{revtex4}
\usepackage{epstopdf}
\usepackage{amsfonts}
\usepackage{amsmath}
\usepackage{amstext}
\usepackage{amssymb}
\usepackage{esint}
\usepackage{graphicx}
\usepackage{graphics}
\usepackage{float}
\usepackage[unicode=true,pdfusetitle,
 bookmarks=true,bookmarksnumbered=false,bookmarksopen=false,
 breaklinks=false,pdfborder={0 0 1},colorlinks=false]{hyperref}
\usepackage{babel}

\hypersetup{colorlinks,linkcolor=blue,citecolor=blue}

\begin{document}

\title{Finite-time Landau-Zener processes and counter-diabatic driving in
open systems: beyond Born, Markov and Rotating-wave approximations}
\author{Zhe Sun}
\email{sunzhe@hznu.edu.cn}
\affiliation{Department of Physics, Hangzhou Normal University, Hangzhou 310036, China}
\affiliation{Department of Physics, National University of Singapore, Singapore 117542}
\author{Longwen Zhou}
\affiliation{Department of Physics, National University of Singapore, Singapore 117542}
\author{Gaoyang Xiao}
\affiliation{Department of Physics, National University of Singapore, Singapore 117542}
\author{Dario Poletti}
\email{dario\_poletti@sutd.edu.sg}
\affiliation{Singapore University of Technology and Design, 8 Somapah Road, 487372
Singapore}
\affiliation{MajuLab, CNRS-UNS-NUS-NTU International Joint Research Unit, UMI 3654,
Singapore}
\author{Jiangbin Gong}
\email{phygj@nus.edu.sg}
\affiliation{Department of Physics, National University of Singapore, Singapore 117542}

\begin{abstract}
We investigate Landau-Zener processes modeled by a two-level quantum system,
with its finite bias energy varied in time and in the presence of a single
broadened cavity mode at zero temperature. By applying the hierarchy
equation method to the Landau-Zener problem, we computationally study the
survival fidelity of adiabatic states without Born, Markov, rotating-wave or
other perturbative approximations. With this treatment it also becomes
possible to investigate cases with very strong system-bath coupling.
Different from a previous study of infinite-time Landau-Zener processes, the
fidelity of the time-evolving state as compared with instantaneous adiabatic
states shows non-monotonic dependence on the system-bath coupling and on the
sweep rate of the bias. We then consider the effect of applying a
counter-diabatic driving field, which is found to be useful in improving the
fidelity only for sufficiently short Landau-Zener processes. Numerically
exact results show that different counter-diabatic driving fields can have
much different robustness against environment effects. Lastly, using a case
study we discuss the possibility of introducing a dynamical decoupling field
in order to eliminate the decoherence effect of the environment and at the
same time to retain the positive role of a counter-diabatic field. Our work
indicates that finite-time Landau-Zener processes with counter-diabatic
driving offer a fruitful test bed to understand controlled adiabatic
processes in open systems.
\end{abstract}

\maketitle

\section{Introduction}

Landau-Zener~(LZ) transitions happen when the bias energy of a quantum
two-level system is swept through an avoided level crossing. For a sweep
with a constant rate from time $t=-\infty $ to $+\infty $, the LZ problem is
exactly solvable and the final survival probability of the adiabatic state
can be obtained asymptotically\thinspace\cite{LZ}. LZ transitions have
attracted considerable interest both theoretically\thinspace and
experimentally\thinspace\cite{Theoretical-1, Theoretical-2,
Theoretical-3,Theoretical-4,experiment2,experiment3}. It has been used to
model dynamical processes in a variety of physical systems\thinspace \cite%
{atoms,BEC}. For instance, with superconducting circuits, which can behave
as controllable quantum two-level systems, LZ and Landau-Zener-St\"{u}%
ckelberg problems were studied by several groups\thinspace\cite%
{super1,super2,super3,super4}. LZ processes controlled by classical or
quantized light fields were also studied in\thinspace\cite{Wubs,Keeling,
Zhe-LZ}.

The influence of environment on quantum systems has motivated numerous
studies on the dissipative LZ problem. Exact results are available at zero
temperature\thinspace\cite{Wubs-Saito-zeroT,Ind-crossing1}, analytical
discussions based on perturbation expansions were derived in\thinspace\cite%
{LZ-dephasing, LZ-Longwen}, and various numerical methods were employed for
the finite-temperature situations\thinspace \cite{Numerical-1, Numerical-2,
Numerical-3, Numerical-4, Numerical-5, Numerical-6, Numerical-7, Numerical-8}%
. The non-monotonic dependence of the LZ transition probability on the sweep
rate was studied in\thinspace\cite{Numerical-1} using a numerically exact
method. Environment parameters, such as temperature, can exponentially
enhance the coherent oscillations generated at a LZ transition\thinspace
\cite{Numerical-3}. Transversal system-bath interactions were found to
influence the survival probability more strongly than longitudinal
system-bath interactions\thinspace \cite{Numerical-5,Numerical-6}.

Put in a more general perspective, the manipulation of instantaneous
eigenstates of a quantum system along its adiabatic paths can be very
useful. In order to avoid nonadiabatic transitions, the operation has to be
executed over a long time scale. Unavoidably, however, a slower process
would allow more unwanted decoherence effects induced by an environment.
Hence, on the one hand a process should be slow such that the system can
stay close to adiabatic states. On the other hand, the process should be
fast so that not too much decoherence can set in before an adiabatic process
ends. This conflict makes it necessary to have a highly reliable treatment
of open quantum systems, which can handle, for example, long-time evolutions
and sometimes strong system-bath couplings.

Usually, the description of the dynamics of open systems starts with a
perturbative theory and involves various approximations, such as the Born,
Markov and rotating-wave approximation~(RWA). An efficient method that
avoids the above approximations was developed by Tanimura \textit{et al}.
\cite{Tanimura1991, Tanimura2010JCP, Tanimura2010PRL, Ishizaki2007,
Tanimura2006}, who established a set of hierarchical equations that includes
all orders of system-bath interactions. The hierarchy equation method has
been successfully employed to describe the quantum dynamics of various
physical, chemical and biophysical systems~\cite{Ishizaki2007, YJ Yan-1, YJ
Yan-2, Ishizaki}. In addition, the hierarchy equation method can be used to
study the dynamics of qubit devices operating at low temperatures\thinspace
\cite{MaJian-Hierarchy, Guo Wei-Berry phase}, when the environment is
usually modelled by a Lorentz-broadened cavity mode. For these reasons, we
shall use the hierarchy equation method to exactly study LZ processes under
the influence of an environment. In particular, we focus on finite-time LZ
processes coupled to an environment. We shall also vary the system-bath
coupling strength from weak to arbitrarily strong regimes, observing a
non-monotonic dependence of the survival fidelity on the coupling strength.
This result indicates that a finite-time LZ process can behave much
differently from the more idealized LZ process that starts and ends with an
infinite energy bias~\cite{Wubs-Saito-zeroT}.

To shorten the time scale of a useful quantum adiabatic operation, various
protocols have been proposed to suppress the nonadiabatic transitions with
additional control fields. Such protocols are often called
counter-diabatic~(CD) driving, shortcuts to adiabaticity, or transitionless
driving~\cite{CD-1,CD-Rice-1,CD-Rice-2,CD-Rice-3,CD-2 Berry, CD-4, CD-5,
CD-6, CD-7}. Following Ref.~\cite{CD-Rice-1,CD-Rice-2,CD-Rice-3}, we refer
to all such quantum control strategies as CD drivings. Some of them have
been implemented in harmonic systems\thinspace \cite{CD-Harmonic-1}, atom
transport\thinspace \cite{CD-atom-trans}, quantum computing\thinspace \cite%
{CD-quant-comp}, quantum simulations\thinspace \cite{CD-quant-simu},
many-body state engineering\thinspace \cite{CD-manybody}, cooling of
nanomechanical resonators\thinspace \cite{CD-cooling}, and so on. CD
drivings have also been suggested as a way to improve the performance of
quantum heat engines~\cite{DelCampoP2014}, although with due
attention to the cost of using them~\cite{ZhengPoletti2015}. Recently, CD
driving was also considered in open systems \thinspace \cite%
{CD-open-1,CD-open-4}. However, these studies were based on a simplified
treatment of the environment.

In this work we investigate the effectiveness of different CD driving
protocols in boosting the fidelity values of LZ processes in the presence of
an environment.
Our investigation is the first of this kind without making any
approximations. From our numerically exact results, it is found that not all
CD driving fields can perform well in the presence of decoherence. While in
the fast driving limit (the associated CD driving field will have extremely
large peak values) all CD driving fields can expectedly do a good job, in
the cases with intermediate sweep rates, the addition of a CD driving field
may even degrade the fidelity. The specific implementation of a CD driving
field may also have a large impact on the fidelity. These results clearly
show that if the total time scale of an adiabatic process cannot be made
very short (e.g., due to hardware implementations or the bandwidth limit of
the CD driving), then the introduction of an additional control field, whose
aim is mainly to suppress the system-environment coupling, can still be of
interest. In the literature a control field capable of modulating and hence
suppressing the system-bath coupling is often called a dynamical
decoupling~(DD) field~\cite{viola1,viola2}. Using a case study at the end of
this work, we indeed discuss the possibility of combining a CD driving field
with a DD field together.



This paper is organized as follows: in Sec.\thinspace II we introduce the
model system and the hierarchy equation method. In Sec.\thinspace III we
show the numerical results of our finite-time LZ dynamics under the
influence of an environment. In Sec.\thinspace IV, two types of CD drivings
are considered and compared. In Sec.\thinspace V, we discuss and numerically
study a possible combination of CD driving and a continuous DD field.
Finally, Sec.\thinspace VI concludes this paper.

\section{System-environment model and hierarchy equation method}

\subsection{ Models, parameters and initial conditions}

We consider a driven two-level system interacting with a bosonic bath. The
total Hamiltonian is ($\hbar =1$ is assumed throughout)
\begin{equation}
H_{\text{tot}}=H_{\text{S}}\left( t\right) +H_{\text{B}}+H_{\text{SB}},
\end{equation}%
where the system-Hamiltonian is chosen to be of LZ-type, i.e., $H_{\text{S}%
}=H_{\text{LZ}}$, with
\begin{equation}
H_{\text{LZ}}=Z\left( t\right) \frac{\sigma _{z}}{2}+X\frac{\sigma _{x}}{2},
\label{H_LZ}
\end{equation}%
where $\sigma _{x,y,z}$ are Pauli matrices. A positive and time-independent
constant $X$ is employed to characterize the interaction between the two
diabatic states $\left\vert \uparrow \right\rangle $ and $\left\vert
\downarrow \right\rangle $ (the eigenstates of $\sigma _{z}$). We assume the
following time-dependent parameter
\begin{equation}
Z\left( t\right) =\frac{\left( z_{f}-z_{0}\right) }{t_{f}}t+z_{0},
\label{Zt}
\end{equation}%
which describes a linearly varying bias-energy between the diabatic states,
with $Z\left( t\right) \in \lbrack z_{0},z_{f}]$ (let $z_{0}=-z_{f}$ and $%
z_{0}<0$ in this paper) and $t\in \lbrack 0,t_{f}]$. The duration of a LZ
process is hence $t_{f}$ and the sweeping rate becomes $v=\left(
z_{f}-z_{0}\right) /t_{f}$. In our study, the boundary of the bias-energy,
determined by $z_{0\text{ }}$and $z_{f}$, is independent of $t_{f}$. By
tuning $t_{f}$, we can study the cases of different sweeping rates. Here,
one should note the difference between the form of Eq.\thinspace (\ref{Zt})
and the conventional LZ driving, i.e., $Z(t)=vt$. In the latter, changing
the evolution time interval is equivalent to changing the boundary of the
bias-energy. Note also that in pure theoretical considerations, an infinite
time interval, i.e., $t\in \left( -\infty ,\infty \right) $ is often chosen
because it allows greater analytical insight, however this implies an
infinite energy bias and is not adopted here.

Now let us briefly recall several results about the standard LZ problem
described by the Hamiltonian in Eq.\thinspace (\ref{H_LZ}). For $X=0$ the
diabatic states $\left\vert \uparrow \right\rangle $ and $\left\vert
\downarrow \right\rangle $ with bias-energies $\pm Z(t)/2$ cross at $%
t=t_{f}/2$ linearly. When $X\neq 0$, the diabatic states are not eigenstates
of the Hamiltonian in Eq.\thinspace (\ref{H_LZ}), and an avoided-level
crossing appears between the adiabatic energy-levels $E_{\pm }\left(
t\right) =\pm \sqrt{\left[ Z(t)\right] ^{2}+X^{2}}/2$ at $t=t_{f}/2$. Thus,
in general, some population transfer occurs. If the bias energy $Z(t)$ is
swept from a large negative value to a large positive value, the diabatic
states coincide with the initial and final adiabatic states. Then it is
possible to compute the survival probability of the qubit ending up in the
initial adiabatic state of $H_{\text{LZ}}$, which is $P_{\text{LZ}}=1-\exp %
\left[ -\pi X^{2}/(2v)\right] $. In the adiabatic limit $X^{2}/v\gg 1$, $P_{%
\text{LZ}}$ will saturate at $1$.

To consider a LZ process in the presence of an environment, we model a
bosonic environment by the bath Hamiltonian $H_{\text{B}}$ and the
system-bath coupling by Hamiltonian $H_{\text{SB}}$,
\begin{eqnarray}
H_{\text{B}} &=&\sum_{k}\omega _{k}b_{k}^{\dagger }b_{k}, \\
H_{\text{SB}} &=&V\sum_{k}\left( g_{k}b_{k}+g_{k}^{\ast }b_{k}^{\dagger
}\right) ,
\end{eqnarray}%
where $\omega _{k}$ indicates the frequency of the $k$-th mode of the bath. $%
b_{k}^{\dagger }$ and $b_{k}$ are the creation and annihilation operators of
the bath. $V=g_{z}\sigma _{z}+g_{x}\sigma _{x}$ denotes the system operators
and it includes longitudinal ($\sigma _{z}$) and transversal ($\sigma _{x}$)
couplings with the mixing coefficient $g_{z}$ and $g_{x}$, respectively. $%
g_{k}$ denotes the individual coupling strength between the qubit and the $k$%
-th mode of the bath. In the following sections, we focus on the transversal
coupling by setting $g_{z}=0$, $g_{x}=1/2$. This is because transversal
interactions in general have stronger influence on population transitions
than longitudinal interactions at the same coupling strength\thinspace \cite%
{Wubs-Saito-zeroT,Numerical-5,Numerical-6}.

The environment is chosen to simulate a single mode of a leaking cavity. Due
to the imperfection of the cavity, the single mode is broadened into a
continuous distribution with a center frequency $\omega_{c}$, and the
qubit-cavity coupling spectrum is usually of a Lorentz-type\thinspace \cite%
{Lorentz-quant-zero, MaJian-Hierarchy}, i.e.,
\begin{equation}
J\left( \omega \right) =\frac{1}{\pi }\frac{\gamma \lambda }{\left( \omega
-\omega _{c}\right) ^{2}+\lambda ^{2}},  \label{Lorentz}
\end{equation}
where $\lambda $ characterizes the broadening width of the cavity mode
(i.e., damping rate), which is connected to the bath correlation time $%
\tau_{B}\sim \lambda ^{-1}$. The perfect-cavity limit is obtained for $%
\lambda\rightarrow 0$, which results in $J\left( \omega \right) =\gamma
\delta \left(\omega -\omega _{c}\right) $. The parameter $\gamma $ reflects
the overall system-bath coupling strength.%

The total system is assumed to start from a product state, i.e.,
\begin{equation}
\rho _{\text{tot}}\left( 0\right) =\rho _{S}\left( 0\right) \otimes \rho
_{B}\left( 0\right) ,
\end{equation}%
where the initial state of the system can be chosen as the ground state of $%
H_{\text{LZ}}$ in Eq.\thinspace (\ref{H_LZ}) at $t=0$. That is, the initial
system density matrix is given by
\begin{equation}
\rho _{S}\left( 0\right) =\left\vert \psi _{g}\left( 0\right) \right\rangle
\left\langle \psi _{g}\left( 0\right) \right\vert ,  \label{rho_s}
\end{equation}%
where $\left\vert \psi _{g}\left( t\right) \right\rangle =-\sin \frac{\theta
_{t}}{2}\left\vert \uparrow \right\rangle +\cos \frac{\theta _{t}}{2}%
\left\vert \downarrow \right\rangle $ is the adiabatic ground state with the
time-dependent superposition coefficient $\theta _{t}=\text{arccot}\left[
Z\left( t\right) /X\right] $. Let $t=0$, one will obtain the initial ground
state corresponding to $\theta _{0}=\text{arccot}\left[ z_{0}/X\right] $.
The adiabatic excited state is $\left\vert \psi _{e}\left( t\right)
\right\rangle =\cos \frac{\theta _{t}}{2}\left\vert \uparrow \right\rangle
+\sin \frac{\theta _{t}}{2}\left\vert \downarrow \right\rangle $. The cavity
mode is assumed to be at zero temperature and starts from a vacuum state $%
\left\vert \psi _{B}\left( 0\right) \right\rangle =\otimes _{k}|0\rangle
_{k} $. The initial bath density matrix is hence given by $\rho _{B}\left(
0\right) =\left\vert \psi _{B}\left( 0\right) \right\rangle \left\langle
\psi _{B}\left( 0\right) \right\vert $.

\subsection{Hierarchy equation method}

For the product form of the initial system-bath state $\rho _{\text{tot}%
}\left( 0\right) =\rho _{S}\left( 0\right) \otimes \rho _{B}\left( 0\right) $%
, the exact dynamics of the system in the interaction picture can be derived
as \cite{Tanimura1991,Tanimura2010PRL}

\begin{eqnarray}
\rho _{S}^{(I)}\left( t\right) &=&\mathcal{\hat{T}}\exp
\{-\int_{0}^{t}\!\!\!dt_{2}\!\int_{0}^{t_{2}}\!\!\!\!\!dt_{1}V^{\times
}\!\left( t_{2}\right) [C^{R}\left( t_{2}-t_{1}\right) V^{\times }\!\!\left(
t_{1}\right)  \notag \\
&&+iC^{I}\left( t_{2}-t_{1}\right) V^{\circ }\left( t_{1}\right) ]\}\rho
_{S}\left( 0\right) ,  \label{rhot_general}
\end{eqnarray}%
where $\mathcal{\hat{T}}$ is the chronological time-ordering operator, which
orders the operators inside the integral such that the time arguments
increase from right to left. Two superoperators are introduced, $A^{\times
}B\equiv \left[ A,\,B\right] =AB-BA$ and $A^{\circ }B\equiv \left\{
A,\,B\right\} =AB+BA$, which simplifies the dynamical equation as well as
the following derivation of the hierarchy equations. In above $C^{R}\left(
t_{2}-t_{1}\right) $ and $C^{I}\left( t_{2}-t_{1}\right) $ are the real and
imaginary parts of the bath time-correlation function
\begin{eqnarray}
C\left( t_{2}-t_{1}\right) &\equiv &\text{Tr}\left[ B\left( t_{2}\right)
B\left( t_{1}\right) \rho _{B}\left( 0\right) \right]  \notag \\
&=&\gamma \exp \left[ -\left( \lambda +i\omega _{c}\right) |t_{2}-t_{1}|%
\right] ,  \label{eq:Lor_corr}
\end{eqnarray}%
respectively, and 
\begin{equation}
B\left( t\right) =\sum_{k}\left( g_{k}b_{k}e^{-i\omega _{k}t}+g_{k}^{\ast
}b_{k}^{\dagger }e^{i\omega _{k}t}\right) .  \label{eq:bath_corr}
\end{equation}%
Note that in Eq.\thinspace (\ref{eq:Lor_corr}) the bath correlation is
already assumed to be an exponential form, a form required to derive the
hierarchy equations. In the single-mode limit $\lambda \rightarrow 0$, we
have $C\left( t_{2}-t_{1}\right) \rightarrow \gamma \exp \left( -i\omega
_{c}\left\vert t_{2}-t_{1}\right\vert \right) $.

To present the hierarchy equations (for completeness) in a convenient form,
we further write the real and imaginary parts of the time-correlation
function (\ref{eq:Lor_corr}) as
\begin{equation}
C^{R}\left( t\right) =\sum_{k=1}^{2}\frac{\gamma }{2}e^{-\nu
_{k}t},\;C^{I}\left( t\right) =\sum_{k=1}^{2}\left( -1\right) ^{k}\frac{%
\gamma }{2i}e^{-\nu _{k}t},
\end{equation}%
where $\nu _{k}=\lambda +(-1)^{k}i\omega _{0}$. Then, following the
procedures shown in \cite{Tanimura1991,Tanimura2010PRL}, the hierarchy
equations of the qubit are found to be
\begin{eqnarray}
\frac{\partial }{\partial t}\varrho _{\vec{n}}\left( t\right) &=&-\left(
iH_{S}^{\times }+\vec{n}\cdot \vec{\nu}\right) \varrho _{\vec{n}}\left(
t\right) -i\sum_{k=1}^{2}V^{\times }\varrho _{\vec{n}+\vec{e}_{k}}\left(
t\right)  \notag \\
&-&i\frac{\gamma }{2}\sum_{k=1}^{2}n_{k}\left[ V^{\times }+\left( -1\right)
^{k}V^{\circ }\right] \varrho _{\vec{n}-\vec{e}_{k}}\left( t\right) ,
\label{eq:hier_eq}
\end{eqnarray}%
where the subscript $\vec{n}=\left( n_{1},\,n_{2}\right) $ is a
two-dimensional index, with integer numbers $n_{1\left( 2\right) }\geq 0$,
and $\rho _{S}\left( t\right) \equiv \varrho _{\left( 0,0\right) }\left(
t\right) $. The vectors $\vec{e}_{1}=\left( 1,\,0\right) $, $\vec{e}%
_{2}=\left( 0,\,1\right) $, and $\vec{\nu}=\left( \nu _{1},\,\nu _{2}\right)
=\left( \lambda -i\omega _{c},\,\lambda +i\omega _{c}\right) $. We emphasize
that $\varrho _{\vec{n}}\left( t\right) $ with $\vec{n}\neq
\left(0,\,0\right) $ are auxiliary operators introduced only for the sake of
computing, they are not density matrices, and are all set to be zero
initially. The hierarchy equations are a set of linear differential
equations, and can thus be solved using, for example, the Runge-Kutta
method. The contributions of the bath to the dynamics of the system,
including both dissipation and Lamb shift, are fully contained in the
hierarchy equation (\ref{eq:hier_eq}).

For numerical computations, the hierarchy equations (\ref{eq:hier_eq}) must
be truncated for a large enough $\vec{n}$. We can increase the hierarchy
order $\vec{n}$ until the results of $\rho _{S}(t)$ converge. The terminator
of the hierarchy equation is
\begin{eqnarray}
\frac{\partial }{\partial t}\varrho _{\vec{N}}\left( t\right) &=&-\left(
iH_{S}^{\times }+\vec{N}\cdot \vec{\nu}\right) \varrho _{\vec{n}}\left(
t\right)  \notag  \label{eq:hier_trunc} \\
&-&i\frac{\gamma }{2}\sum_{k=1}^{2}n_{k}\left[ V^{\times }+\left( -1\right)
^{k}V^{\circ }\right] \varrho _{\vec{N}-\vec{e}_{k}}\left( t\right) ,  \notag
\\
&&
\end{eqnarray}%
where we dropped the deeper auxiliary operators $\varrho _{\vec{N}+\vec{e}%
_{k}}$. The numerical results in this paper were all converged, and the
density matrix $\rho _{S}(t)$ was tested to be positive. The detailed
derivation of Eq.\thinspace (\ref{eq:hier_eq}) can be found in
Refs.\thinspace \cite{Tanimura1991,MaJian-Hierarchy}.

\section{Finite-time Landau-Zener processes in an envrionment}

\begin{figure}[tbp]
\includegraphics[width=0.56\textwidth, height=0.4\textwidth]{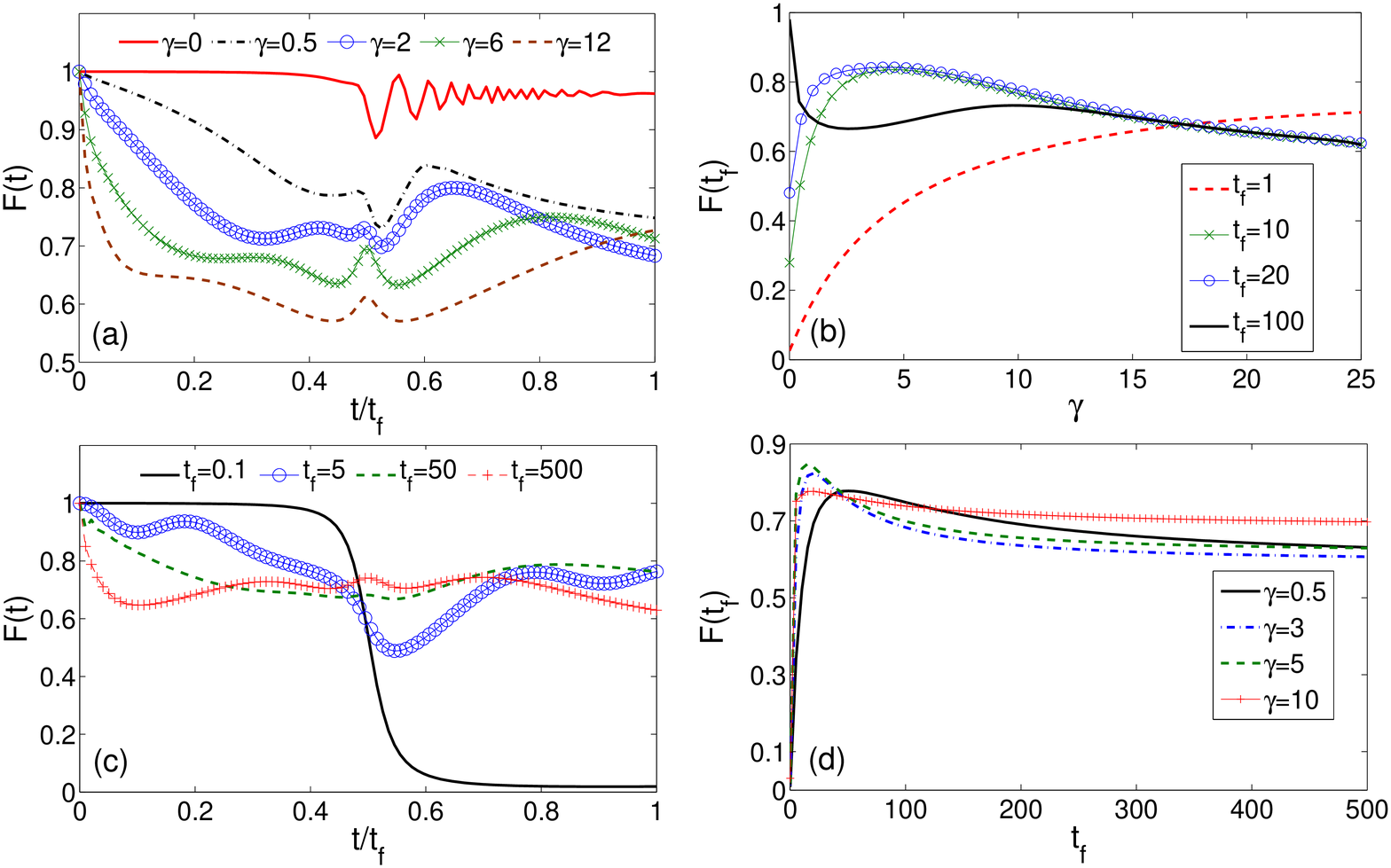}
\caption{(Color online) \textbf{(a)} The fidelity with instantaneous
adiabatic states, defined as $F(t)=\left\langle \protect\psi_g(t)|\protect%
\rho(t)|\protect\psi_g(t)\right\rangle $, versus scaled time $t/t_f$ for
different values of the system-bath coupling strength parameter $\protect%
\gamma$. The duration of the finite-time LZ processes is chosen to be $%
t_f=100$. (b) Final fidelity, defined as $F(t_f)=\left\langle \protect\psi%
_g(t_f)|\protect\rho(t_f)| \protect\psi_g(t_f)\right\rangle $, versus $%
\protect\gamma$ for different durations of the LZ processes. (c) The
fidelity with instantaneous adiabatic states versus scaled time $t/t_f$ for
different values of $t_f$, with $\protect\gamma=5$. (d) Final fidelity
versus $t_f$ for different values of $\protect\gamma$. Other system and
environment parameters are given by $z_f=-z_0=6$, $X=0.5$ [see Eq.~(\protect
\ref{H_LZ})] and $\protect\omega_c=0.5$ [see Eq.~(\protect\ref{Lorentz})].
As explained in the main text, all plotted quantities in this work are
dimensionless.}
\end{figure}

We are now ready to investigate, using numerically exact results, the
features of finite-time LZ processes in an environment modeled above. Some
parameters are fixed throughout, i.e., for an arbitrary energy unit, we
choose the scaled values $X=0.5$, $z_{f}=-z_{0}=6$, $\lambda =0.5$ and $%
\omega _{c}=0.5$, where a relatively large initial and final bias-energy is
chosen, i.e., $\left\vert z_{0}\right\vert=\left\vert z_{f}\right\vert =12X$%
, and the central frequency of the bath is assumed to be equal to the
minimal energy gap of the adiabatic states, i.e., $\omega _{c}=X$.


In conventional physical systems, the intrinsic energy scale of the system
is larger than the energy-scale of the system-bath coupling. However, in
artificial quantum devices\thinspace \cite{super1,super2}, this may not be
the case. Therefore, we consider both weak and strong coupling cases and
this is possible because the hierarchy equation method is not perturbative.

One central quantity in our investigation is $F\left( t\right) =\left\vert
\left\langle \psi _{g}\left( t\right) \right\vert \rho \left( t\right)
\left\vert \psi _{g}\left( t\right) \right\rangle \right\vert $, which
describes the overlap between the actual time-evolving state and the target
instantaneous ground state. In Fig.\thinspace 1(a), we show $F\left(
t\right) $ versus the scaled time $t/t_{f}$ for different values of $\gamma $%
. The (red) solid line displays the case without environment ($\gamma =0$),
wherein a high final fidelity is found. This high fidelity can be attributed
to the slow sweeping rate of $v=0.12$ ($t_{f}=100$), which gives a large
ratio of $X^{2}/v\approx 2.1$. Indeed, if the final fidelity approximately
obeys the standard LZ probability formula $F\left( t_{f}\right) \approx
1-\exp \left[ -\pi X^{2}/(2v)\right] $ [note, strictly speaking, this
formula is only for an infinite-time LZ with $t\in (-\infty ,\infty )$], one
expects the final fidelity close to unity. However for $\gamma \neq 0$ the
fidelity is strongly affected. Before the LZ transition point at $%
t/t_{f}=1/2 $, an increase in the system-bath coupling strength $\gamma $
leads to a decrease in the fidelity. After the LZ transition point, a larger
$\gamma $ can result in increased fidelity, i.e., there exists a
non-monotonic dependence of the final fidelity on the coupling strength $%
\gamma $.

This interesting nonmonotonic behavior is shown even more clearly in
Fig.\thinspace 1(b), where we plot the final fidelity $F\left( t_{f}\right) $
versus $\gamma $ for different durations of a LZ process. The (black) solid
line describes a case with a slow sweep rate of $v=0.12$ ($t_{f}=100$),
wherein the LZ time scale\thinspace \cite{LZ}, i.e., $\tau _{\text{LZ}}\sim
\max [X/v,1/\sqrt{v}]$ is larger than the bath correlation time $\tau
_{B}\sim 1/\lambda $. For this case there is enough time for the environment
to influence the LZ process. With the increase of $\gamma $, we find first a
local minimum of the final fidelity $F\left( t_{f}\right) $, then a peak,
and finally a gradual decrease. Next we consider shorter processes with $%
t_{f}=20$ and $10$, i.e., $v=0.6$ and $1.2$, respectively. The associated
time scale $\tau _{\text{LZ}}$ becomes smaller than $\tau _{B}$. In these
two cases, the fidelity starts from low values due to the small ratios of $%
X^{2}/v\approx 0.42$ and $0.21$. Now, as $\gamma $ increases, there only
exists a peak in the final fidelity. Further decreasing the duration to $%
t_{f}=1$ (i.e., $v=12$), the non-monotonic behavior vanishes, so $F\left(
t_{f}\right) $ rises\ monotonically for the entire shown regime of $\gamma $.

Recalling one of the results of Wubs \textit{et al.}\thinspace \cite%
{Wubs-Saito-zeroT}
\begin{equation}
F\left( \infty \right) =1-e^{-\pi W^{2}/(2v)},  \label{Ff-Wubs}
\end{equation}%
where $W^{2}=X^{2}+\gamma $
, and $\gamma =\int d\omega J\left( \omega \right) $ is the effective
coupling strength obtained by integrating the spectral density of the
environment. In Ref.~\cite{Wubs-Saito-zeroT} only a monotonic increase of $%
F\left( t_{f}\right) $ versus system-bath coupling was found regardless of
whether the sweeping rate is fast or slow. A simple check reveals the origin
of the difference. In Ref.~\cite{Wubs-Saito-zeroT} a fundamental requirement
was that the process was conducted for an infinite evolution time, i.e., $%
t\in (-\infty ,\infty )$. This implies that the initial and final energy
bias of the qubit system must be infinitely large as compared with an
arbitrarily strong system-bath coupling.
In contrast, our finite-time LZ process involves situations for which the
system-bath couplings are comparable to the energy bias at the start or in
the end. For this reason, a complicated interplay between non-adiabatic
transitions and environment-induced excitation or de-excitation can be
expected. To confirm this insight, we note that for the studied range of the
system parameters, the nonmonotonic behavior discussed above emerges only
when the sweeping rate yields a time scale that is comparable with that
associated with environment induced transitions.

Similar nonmonotonic phenomena have been found even in an example with pure
dephasing dynamics\thinspace \cite{LZ-dephasing, LZ-Longwen}. There the
authors analytically obtained the final fidelity based on the first-order
adiabatic perturbation theory regarding the sweeping rate $v$. The
approximate result successfully predicts a minimum of $F\left( t_{f}\right) $%
, which qualitatively agrees with our result shown in Fig.\thinspace 1(b)
for a quite slow sweeping with $t_{f}=100$. However, it fails to present the
maximum of $F\left( t_{f}\right) $ shown in Fig.\thinspace 1(b) for stronger
system-bath coupling strength, a situation that is definitely beyond a
pure-dephasing model.

Fig.\thinspace 1(c) shows the fidelity versus scaled time for varying
durations of LZ processes. The (black) solid line describes the case of fast
sweeping ($t_{f}=0.1$), where one finds the final fidelity ending at a very
small value. For the sake of comparison, we increase $t_{f}$ to $5$, $50$,
and $500$. As a result, more complex oscillations of $F\left( t\right) $ are
observed and the fidelity values after the transition point $t/t_{f}=1/2$
are higher than the fast sweeping case with $t_{f}=0.1$. Interestingly, the
computational results show an upper limit of the final fidelity as $t_{f}$
increases. Indeed, the final fidelity for $t_{f}=500$, is not much different
from that with $t_{f}=50$. This is an obvious environment-induced
suppression, which can be seen more clearly in the panel (d), to be
discussed next.

In Fig.\thinspace 1(d), we plot the final fidelity $F\left( t_{f}\right) $
versus $t_{f}$ for different values of the system-bath coupling strength
parameter $\gamma $. There one can clearly observe the non-monotonic
dependence of fidelity on $t_f$. For short LZ processes, the final fidelity
values are small and different system-bath couplings can hardly induce
significant differences. With increased $t_{f}$, different values of $\gamma$
yield different peaks as well as different saturated values of fidelity,
constituting interesting features again that are absent in infinite-time LZ
processes at zero temperature.

\section{LZ processes with counter-diabatic driving and an environment}

\begin{figure}[tbp]
\includegraphics[width=0.52\textwidth, height=0.36\textwidth]{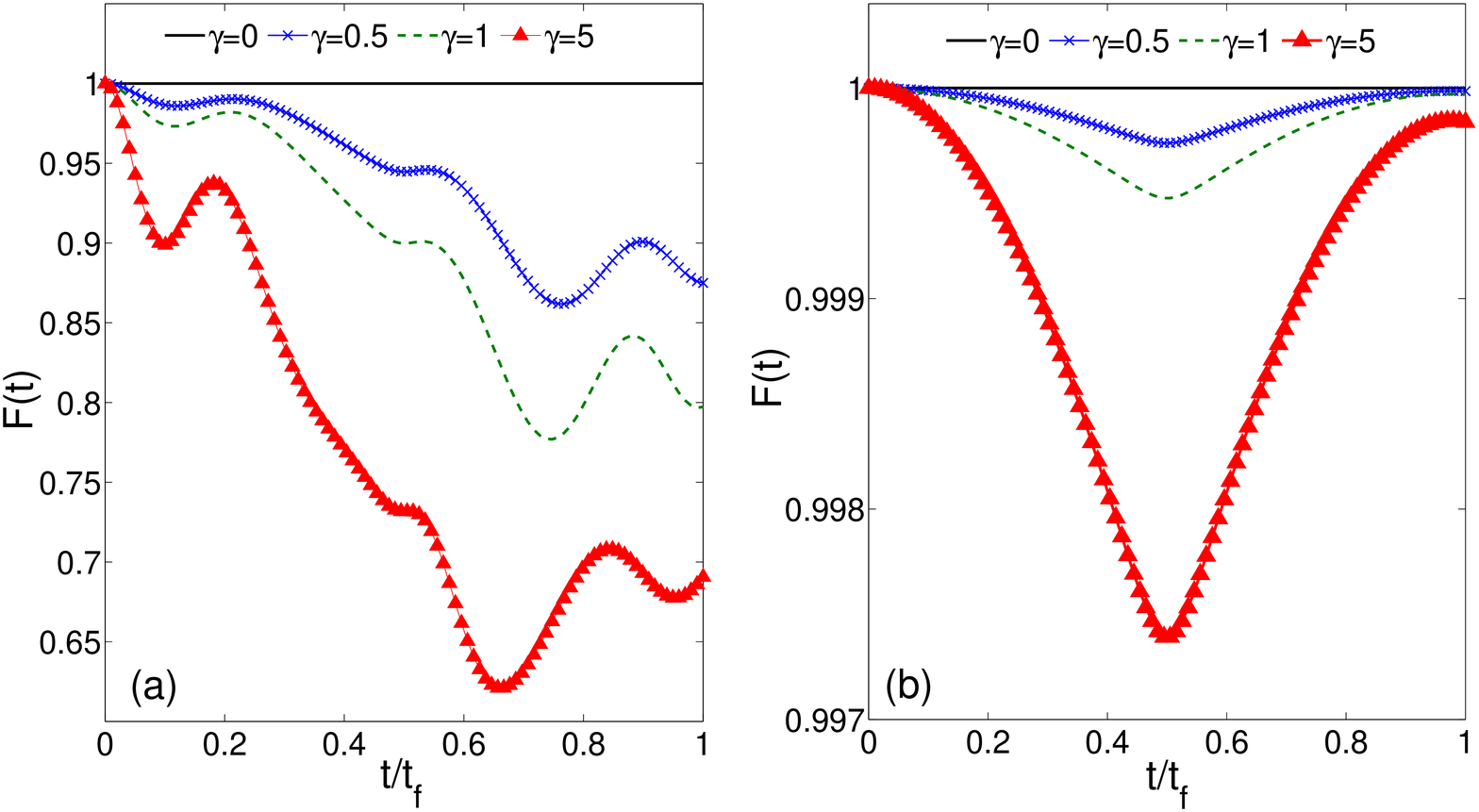}
\caption{(Color online) The fidelity with instantaneous adiabatic states
versus scaled time $t/t_f$, for different system-bath coupling strength in
the presence of CD driving. The duration of the LZ processes is chosen $%
t_f=5 $ in panel (a) and $t_f=0.1$ in panel (b). In both panels, the top
solid (black) lines correspond to the cases without a bath, i.e., $\protect%
\gamma=0 $, for which the CD driving leads to perfect unity fidelity. Other
system and environment parameters are the same as in Fig.~1. Note that panel
(a) and (b) used different scales on the $y$ axis.}
\end{figure}

In this section we aim to study how CD driving might improve the fidelity of
the time-evolving state in an LZ process (as compared with the instantaneous
adiabatic states). We first briefly outline the idea behind the CD driving
strategy and then present our results.

\subsection{Counter-diabatic driving}

A CD\thinspace \cite{CD-Rice-1,CD-2 Berry} driving field is designed to
eliminate all unwanted nonadiabatic transitions. Specifically, here the
reference Hamiltonian is $H_{\text{LZ}}$ that determines the instantaneous
adiabatic states and hence an ideal adiabatic path. One can now construct an
auxiliary driving Hamiltonian $H_{\text{CD}}$ to ensure the following: when
the system's initial state is prepared in the eigenstate of $H_{\text{LZ}}$
at $Z=Z(0)$, i.e., $\left\vert \psi _{n}\left( Z(0)\right) \right\rangle $,
its ensuing time evolution will track the instantaneous eigenstate $%
\left\vert \psi _{n}\left( Z(t)\right) \right\rangle $. To achieve this the
following target unitary operation is required, i.e.,
\begin{equation}
U_{\text{target}}=\sum\limits_{n}e^{i\phi _{n}\left( t\right) }\left\vert
\psi _{n}\left( Z(t)\right) \right\rangle \left\langle \psi _{n}\left(
Z(0)\right) \right\vert ,
\end{equation}%
where $\phi _{n}$ is a phase factor whose time dependence can be arbitrarily
chosen. This unitary evolution can be directly realized by the driving
Hamiltonian $H_{\text{S}}=i\dot{U}_{\text{target}}U_{\text{target}}^{\dagger
}$, i.e.,
\begin{eqnarray}
H_{\text{S}} &=&-\sum\limits_{n}\dot{\phi}_{n}\left\vert \psi _{n}\left(
Z(t)\right) \right\rangle \left\langle \psi _{n}\left( Z(t)\right)
\right\vert  \notag \\
&&+i\sum\limits_{n}\left[ \partial _{t}\left\vert \psi _{n}\left(
Z(t)\right) \right\rangle \right] \left\langle \psi _{n}\left( Z(t)\right)
\right\vert .  \label{H_driving}
\end{eqnarray}%
The first term in the above equation is in diagonal form, and so it always
commutes with $H_{\text{LZ}}$. Due to the freedom in the choice of the phase
$\phi _{n}\left( t\right) $ in Eq.\thinspace (\ref{H_driving}), this first
term may take many different forms. A typical choice is to make $\phi _{n}$
identical with the dynamical phase associated with the ideal adiabatic
process naturally generated by $H_{\text{LZ}}$. That is, $\phi
_{n}=-\int\nolimits_{0}^{t}E_{n}[Z(\tau )]d\tau $, where $E_{n}[Z(t)]$ is
the instantaneous eigenvalue of the adiabatic state $\left\vert \psi
_{n}\left( Z(t)\right) \right\rangle $. Under this choice the first term in
Eq.~(\ref{H_driving}) is nothing but $H_{\text{LZ}}$ itself. The second term
in the above equation is often called the CD driving field, i.e.,
\begin{equation}
H_{\text{CD}}=i\sum\limits_{n}\left[ \partial _{t}\left\vert \psi
_{n}(Z\left( t\right) )\right\rangle \right] \left\langle \psi _{n}(Z\left(
t\right) )\right\vert .
\end{equation}%
Using the specific form of $|\psi _{n}\left( Z(t)\right) \rangle $ for our
LZ system, one finds
\begin{equation}
H_{\text{CD}}=\frac{\dot{\theta}_{t}}{2}\sigma _{y},  \label{CD-key}
\end{equation}%
where $\theta _{t}=\text{arccot}\left[ Z\left( t\right) /X\right] $. The
system Hamiltonian hence becomes $H_{\text{S}}=H_{\text{LZ}}+H_{\text{CD}}$.

\begin{figure}[tbp]
\includegraphics[width=0.52\textwidth, height=0.37\textwidth]{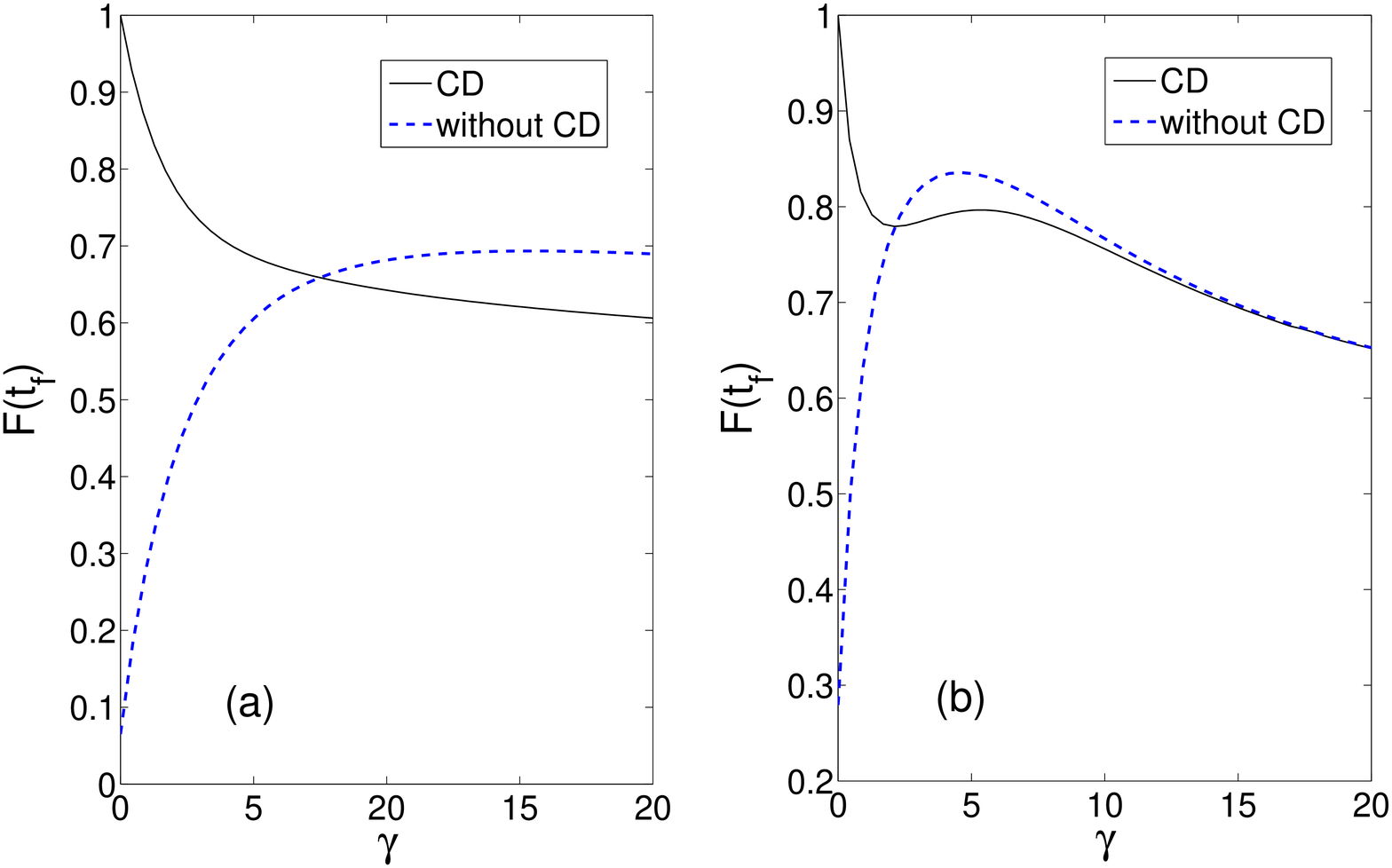}
\caption{(Color online) The final fidelity values versus the system-bath
coupling strength parameter $\protect\gamma$, with the total duration of LZ
processes given by $t_f=2$ in \textbf{(a)} and $t_f=10$ in \textbf{(b)},
with and without CD driving. Other system and environment parameters are the
same as in Fig.~1.}
\end{figure}

From the derivation above, it can be noted that the component $H_{\text{LZ}}$
in $H_{\text{S}}$ only provides a dynamical phase during the designed
evolution. Different choices of $\phi_{n}$ will produce different diagonal
operations in the eigenspace spanned by $\left\vert \psi _{n}\left(
Z(t)\right) \right\rangle $. The actual component that is responsible for
suppressing non-adiabatic transitions is the $H_{\text{CD}}$ component. For
example, which we will make use of later, if we set $\phi _{n}=0$, then $H_{%
\text{S}}=H_{\text{CD}}$ is equally competent to give rise to an adiabatic
evolution path that tracks perfectly the instantaneous adiabatic states $%
\left\vert \psi _{n}\left( Z(t)\right) \right\rangle $.

The above design strategy does not have any environment in the picture. The
effectiveness of CD driving in the presence of an environment is hence of
considerable interest. We now present our computational results on this
issue. In Fig.\thinspace 2, the total system Hamiltonian is chosen as $H_{%
\text{S}}=H_{\text{LZ}}+H_{\text{CD}}$. System and environment parameters
such as $z_{0}$, $z_{f}$, $X$, $\omega _{c}$, $\lambda $ are the same as
those in Fig.\thinspace 1. We calculate the fidelity versus scaled time $%
t/t_{f}$ for different values of $\gamma$. Panel (a) of Fig.\thinspace 2
shows three cases with $\gamma$ increasing from $0.5$, $1$ to $5$, all with $%
t_{f}=5$, a process not too short or too long. It is found that the fidelity
$F\left( t\right) $ oscillates with time and becomes much lower than unity
in the end. Let us now compare these results with those in Fig.\thinspace
1(c) also with $\gamma =5$ and $t_{f}=5$, that is, compare the (blue) solid
line (connected with circles) in Fig.\thinspace 1(c) and the (red) solid
line (connected with triangles) in Fig.\thinspace 2(a). Remarkably, it is
found that the final fidelity with CD driving is even lower than that
without CD driving. In other words, the CD driving, which is designed in a
close-system set up, may unfortunately favor environment-induced effects.

The results shown in Fig.\thinspace 2(b) are for very short LZ processes,
with $t_{f}=0.1$ or $v=120$. In this case, the driving time scale $1/\sqrt{v}%
\ll \tau _{B}$. Intuitively then, the environment has virtually no time to
take effect. Indeed, for all the three cases shown in Fig.\thinspace 2(b),
the fidelity always stays extremely close to unity, even for $\gamma =5$
that represents a case with very strong system-environment coupling. By
contrast, the bare $H_{\text{LZ}}$ Hamiltonian without CD driving only
yields fairly low fidelity values [see the black solid line in
Fig.\thinspace 1(c)]. An obvious conclusion is that the CD driving can work
well only for fast sweeping cases. However, one should note that to realize
very fast sweeping poses a challenge in actual implementations.

Fig.~3 depicts the final fidelity values as a function of the
system-environment coupling strength $\gamma$, with or without CD driving.
We choose two intermediate values of the sweeping rate to make an
interesting comparison, with $t_f=2$ in panel (a) and $t_f=10$ in panel (b).
In both panels, a crossing of the two fidelity curves occurs, which again
demonstrates the existence of a regime that the CD driving leads to a lower
fidelity than that without CD. Further calculations (not shown) suggest that
for the present choice of all other parameters, this crossing occurs for $%
0.7<t_{f}<40$. Beyond this intermediate regime, i.e., for very fast
sweeping, CD driving is always beneficial (see Fig.~2b), whereas for very
slow sweeping, switching on CD driving does not cause a difference to the
final fidelity values.

\subsection{A variant of counter-diabatic driving}

\begin{figure}[tbp]
\includegraphics[width=0.53\textwidth, height=0.38\textwidth]{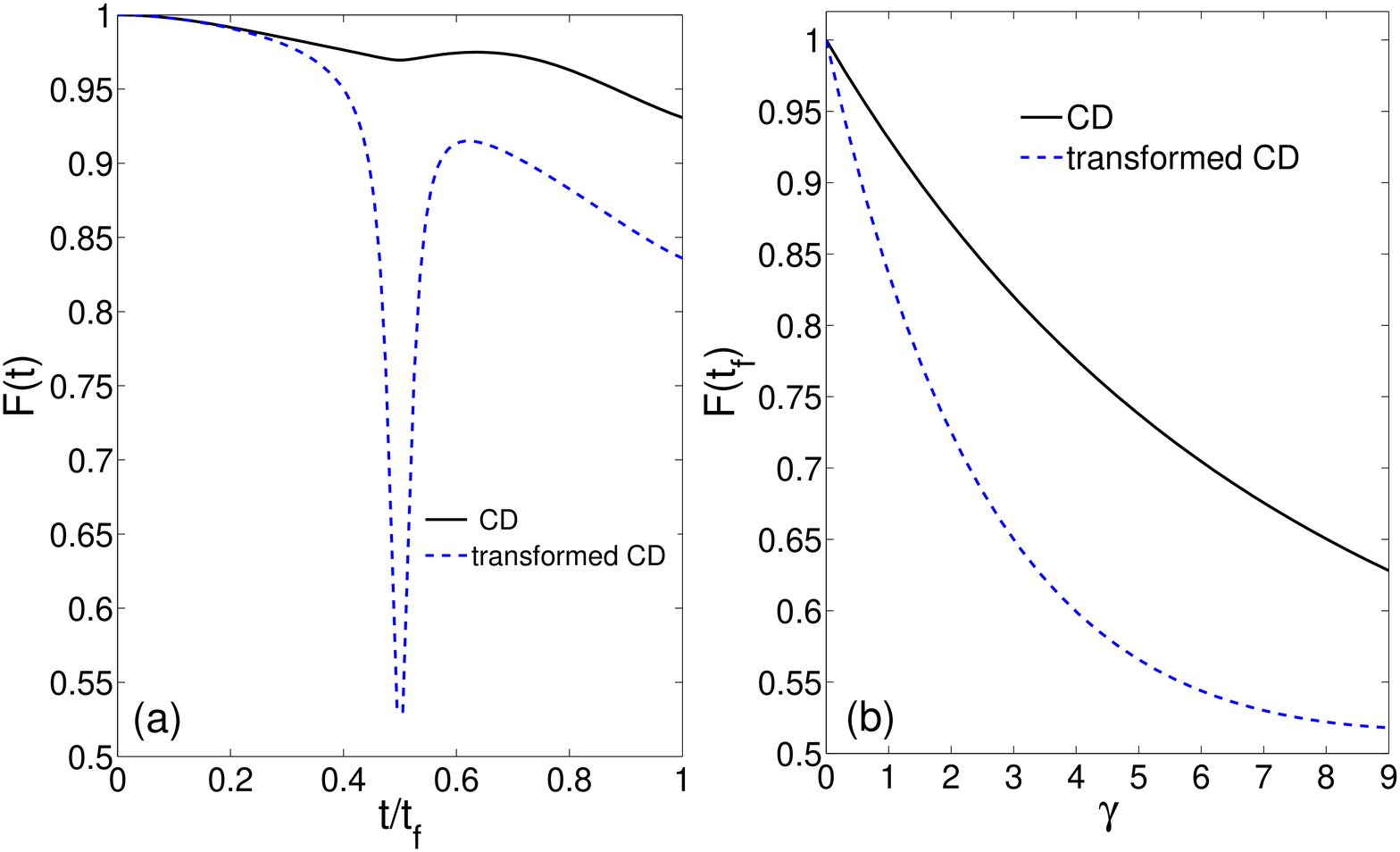}
\caption{(Color online) (a) The fidelity with instantaneous adiabatic states
versus scaled time $t/t_f$, with the system-bath coupling strength parameter
$\protect\gamma=1.0$, for CD driving and transformed CD driving. The
duration of the LZ processes is chosen to be $t_f=1$. (b) Final fidelity
values for CD driving and transformed CD driving, as a function of $\protect%
\gamma$, with $t_f=1$. Other system and environment parameters are the same
as in Fig.~1.}
\end{figure}

In this subsection we study the performance of a variant of CD driving \cite%
{CD-7}. The idea behind transformed CD driving is as follows. If the main
concern is to reach a final adiabatic state with high fidelity from a given
initial adiabatic state, then there is no need to impose conditions on the
evolution operator at intermediate times. Rather, only the unitary evolution
operator associated with the whole quantum process is under consideration
for designing a control field. One motivation to design such CD variants is
to avoid the potential difficulties in the actual experimental realizations.
Indeed, by considering different types of transformations from the adiabatic
path of the instantaneous eigenstates of $H_{\text{LZ}}$ to other paths in
the Hilbert space, one obtains different driving fields to execute possible
variants of CD, which we call transformed CD in this work.

In the present LZ problem, one transformed CD driving was introduced by del
Campo\thinspace \cite{CD-manybody}. It requires $Z\left( t\right)
=z_{0}+6\delta s^{5}-15\delta s^{4}+10\delta s^{3}$, with $s=t/t_{f}$ and $%
\delta =z_{f}-z_{0}$, with the transformed CD driving Hamiltonian given by
\begin{equation}
H_{\text{TCD}}\left( t\right) =\left[ Z\left( t\right) -\dot{\eta}%
\left(t\right) \right] \frac{\sigma _{z}}{2}+P\left( t\right) \frac{\sigma
_{x}}{2},  \label{CD_transformed}
\end{equation}
where $P=\sqrt{X^{2}+\dot{\theta}_{t}^{2}}$, \ $\theta _{t}=\text{arccot}%
\left[ Z\left( t\right) /X\right] $ and $\eta \left( t\right) =\arctan\left(
\dot{\theta}_{t}/X\right) $. As seen above, the main interesting feature of
this transformed CD driving is that it does not contain $\sigma _{y}$ and
hence it is more analogous to the original $H_{\text{LZ}}$ Hamiltonian.

It is now of great interest to compare the performance of the transformed CD
driving with that of the above-studied CD driving, in the presence of an
environment. Panel (a) of Fig.\thinspace 4 shows the fidelity with
instantaneous adiabatic states versus scaled time, with $\gamma =1$.
Somewhat expected, before the LZ process is over the fidelity of transformed
CD is systematically smaller. Interestingly, this difference sustains at the
end of the LZ process. That is, the final fidelity of the transformed CD is
not as satisfactory as that of CD. Panel (b) Fig.\thinspace 4 compares the
final fidelity values as a function of $\gamma $. Clearly, for the whole
shown range of system-bath coupling strength, transformed CD is less
effective than CD in reaching the final target state (though in the unitary
limit these two driving strategies give the same final result by
construction).
However, it remains an open question to see if it is possible for a
transformed CD driving to be more robust against an environment than a
default CD driving.

\section{CD driving with continuous dynamical decoupling}

Our results in previous sections clearly show that, so long as the quantum
adiabatic operation is not very short, the presence of an environment poses
a hurdle to realizing perfect adiabatic operations even with CD driving. We
are thus motivated to consider a simple dynamical decoupling (DD) protocol
to suppress the effect of system-bath coupling. This is highly nontrivial
because in general, the introduction of a DD field can easily interfere with
the design of CD driving. This section hence serves as a starting point to
motivate further studies on this interesting problem.

For completeness, we first briefly introduce continuous DD, which is based
on a continuous control Hamiltonian, denoted as $H_{\text{DD}}$, to
effectively reduce system-bath coupling \thinspace \cite{DD-2, DD-3, DD-4,
DD-Adam-Gong}. The total system-bath Hamiltonian is hence given by
\begin{eqnarray}
H_{\text{tot}} &=&H_{\text{S}}+H_{\text{DD}}+H_{\text{B}}+H_{\text{SB}},
\notag \\
&\equiv &H^{\prime }+H_{DD}.
\end{eqnarray}%
A unitary transformation $U_{\text{DD}}$ generated by $H_{\text{DD}}$ alone
can be expressed as
\begin{equation}
U_{\text{DD}}\left( t\right) =\mathit{\hat{T}}\exp \left[ -i\int_{0}^{t}H_{%
\text{DD}}\left( \tau \right) d\tau \right] .
\end{equation}%
In continuous DD, $U_{\text{DD}}\left( t\right) $ satisfies two conditions.
First, $U_{\text{DD}}\left( t\right) $ should be periodic in time with a
period denoted by $t_{D}$, i.e.,%
\begin{equation}
U_{\text{DD}}\left( t+t_{D}\right) =U_{\text{DD}}\left( t\right) .
\label{DD-condition-1}
\end{equation}%
Second,
\begin{equation}
\int_{0}^{t_{D}}U_{\text{DD}}^{\dagger }\left( \tau \right) H_{\text{SB}}U_{%
\text{DD}}\left( \tau \right) d\tau =0.  \label{DD-condition-2}
\end{equation}%
The mechanism of continuous DD can then be easily understood. In particular,
one may use $H_{\text{DD}}$ as a reference Hamiltonian to define an
interaction representation or a rotating frame. In that representation, the
total Hamiltonian becomes $\tilde{H}^{\prime }=U_{DD}^{\dagger }{H}^{\prime
}U_{DD}$. The associated total system-bath propagator becomes
\begin{equation}
\tilde{U}_{\text{tot}}\left( t\right) =\mathit{\hat{T}}\exp \left[
-i\int_{0}^{t}\tilde{H}^{\prime }\left( \tau \right) d\tau \right] .
\end{equation}%
Dividing the total time interval $\left( 0\sim t\right) $ into $N$ pieces of
width $t_{D}=t/N$ ($N$ is a positive integer number) and using the Magnus
expansion to the first order of $1/N$ \thinspace \cite{Magnus-expansion},
one has
\begin{eqnarray}
\tilde{U}_{\text{tot}}\left( t\right) &=&\prod\limits_{n=1}^{N}\mathit{\hat{T%
}}\exp \left[ -i\int_{(n-1)t_{D}}^{nt_{D}}\tilde{H}^{\prime }\left( \tau
\right) d\tau \right]  \notag \\
&\approx &\prod\limits_{n=1}^{N}\exp \left[ -i\int_{(n-1)t_{D}}^{nt_{D}}%
\tilde{H}^{\prime }\left( \tau \right) d\tau \right] .
\end{eqnarray}%
Due to the conditions in Eqs.\thinspace (\ref{DD-condition-1}) and (\ref%
{DD-condition-2}), terms with $H_{\text{SB}}$ will vanish. That is, $%
-i\int_{(n-1)t_{D}}^{nt_{D}}\tilde{H}^{\prime }\left( \tau \right) d\tau
\rightarrow -i\int_{(n-1)t_{D}}^{nt_{D}}[H_{\text{S}}(\tau )+H_{\text{B}%
}\left( \tau \right) ]d\tau $. Then $\tilde{U}_{\text{tot}}\left( t\right) $
will factor into a product of independent evolution operators for the system
and for the bath. Note also that at $t=nt_{D}$, $U_{\text{DD}}=1$ and hence $%
\tilde{U}_{\text{tot}}$ is precisely the overall system-bath evolution
operator in the original representation. That is, in the limit of $%
t_{D}\rightarrow 0$, the system's evolution will completely decouple from
that of the bath, with the evolution operator at $t=nt_{D}$ given by $\tilde{%
U}_{S}\left( t\right) =\mathit{\hat{T}}\exp \left[ -i\int_{0}^{t}U_{\text{DD}%
}^{\dagger }\left( \tau \right) H_{\text{S}}\left( \tau \right) U_{\text{DD}%
}\left( \tau \right) d\tau \right] $

\begin{figure}[tbp]
\includegraphics[width=0.5\textwidth, height=0.4\textwidth]{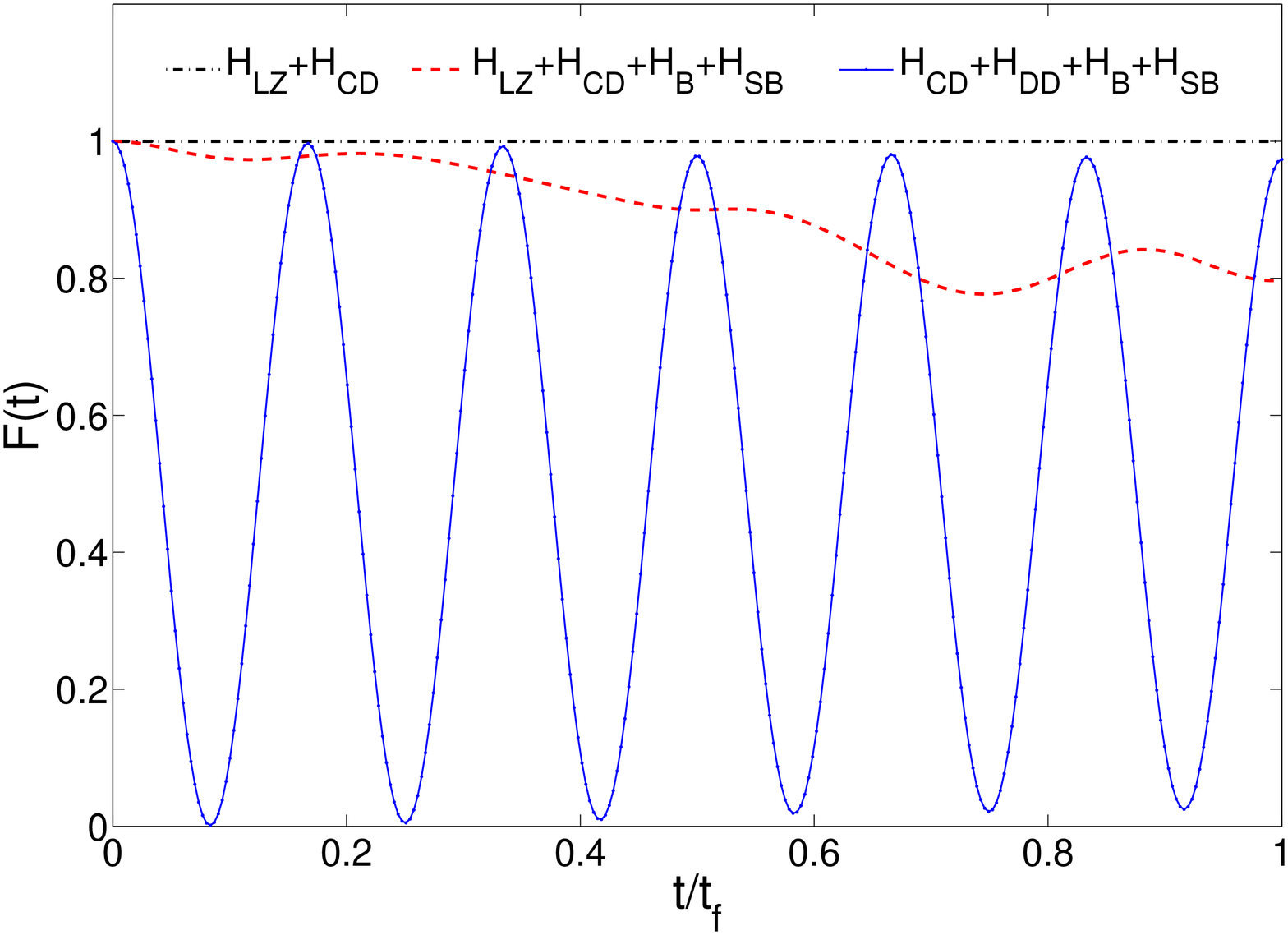}
\caption{(Color online) The fidelity with instantaneous adiabatic states
versus scaled time $t/t_f$, with $t_f=5$ and $\protect\gamma=1$. Other
system and environment parameters are the same as in previous figures.
(black) Dashed-dotted line represents a reference case, namely, CD driving
without an environment, with $H_{\text{S}}=H_{\text{LZ}}+H_{\text{CD}}$.
(red) Dashed line is for CD driving without DD, with the total Hamiltonian
given by $H_{\text{LZ}}+H_{\text{CD}}+H_{\text{SB}}+H_{\text{B}}$. (blue)
Solid line is for a case with ``CD+DD'' fields, with the total Hamiltonian
given by $H_{\text{CD}}+H_{\text{DD}}+H_{\text{SB}}+H_{\text{B}}$.}
\end{figure}

The DD mechanism explained above makes it clear that under a DD field, the
effective system Hamiltonian becomes that of $\tilde{H}_{\text{S}}(t)=U_{%
\text{DD}}^{\dagger }(t) H_{\text{S}}(t)U_{\text{DD}}\left( t \right)$.
Before switching on a DD field, $H_\text{S}(t)$ is meant to implement a CD
driving. But now with DD, can the effective Hamiltonian $\tilde{H}_\text{S}%
(t)$ still serve the purpose of CD driving or its variant? This can be a
complicated question. Indeed, in general, $U_{\text{DD}}$ might introduce
transitions between different adiabatic states. However, this still does not
exclude special solutions, one of which is discussed below.

Because the system-bath coupling is of the $\sigma _{x}$ form, a suitable
and simple choice of $H_{\text{DD}}$ is \cite{DD-2,DD-3,DD-Adam-Gong}
\begin{equation}
H_{\text{DD}}=Y_{D}\sigma _{y}=\frac{\pi }{t_{D}}\sigma _{y},  \label{H_DD}
\end{equation}%
This DD Hamiltonian tends to rapidly rotate the $\sigma _{x}$ coupling and
then average the system-bath coupling to zero. On the other hand, in Sec.~IV
A it is noted that $H_{\text{CD}}=\frac{\dot{\theta}_{t}}{2}\sigma _{y}$
[see Eq.~(\ref{CD-key})] alone suffices to realize an adiabatic path
precisely tracing the instantaneous adiabatic states of $H_{\text{LZ}}$.
That is, the Hamiltonian $H_{\text{LZ}}$ itself, which only introduces a
dynamical phase to the instantaneous adiabatic states, is not essential, and
can hence be excluded, from the perspective of CD driving. Remarkably then,
if $H_{\text{S}}$ is simply chosen as $H_{\text{CD}}$, then both $H_{\text{DD%
}}$ and $H_{\text{S}}$ are of the $\sigma _{y}$ form, and so $\tilde{H}_{%
\text{S}}(t)=U_{\text{DD}}^{\dagger }(t)H_{S}(t)\left( \tau \right) U_{\text{%
DD}}\left( t\right) =H_{\text{S}}(t)$. Therefore, at least in this special
case of CD driving, a DD field is able to effectively decouple the
system-bath coupling and at the same time retain the CD driving.


In Fig.\thinspace 5, we present our computational results for a pure CD\
driving as well as a CD driving plus the above-introduced DD field, in terms
of the fidelity with instantaneous eigenstates of $H_{\text{LZ}}$. As a
reference point, The (black) dashed-dotted line depicts the case without
environment, whose fidelity stays at unity. We then switch on the
system-bath coupling with $\gamma=1$. The (blue) solid line depicts the case
with CD and DD, the oscillations in the fidelity are induced by $H_{\text{DD}%
}$ and high fidelity values are reached periodically. The final fidelity $%
F(t_{f})\approx 0.97 $ is much higher than if only CD driving is used (which
gives $F(t_{f})\approx 0.8$ shown in the red dashed line). We have also
checked that if we let $H_{\text{S}}=H_{\text{LZ}} +H_{\text{CD}}$ to
realize an adiabatic operation in the presence of an environment, then
turning on a DD field as described above will not help to improve the
fidelity.

Theoretically, $t_D=\frac{\pi}{Y_D}$ describes the rotation period of $U_{%
\text{DD}}$. In accord with the above theory of continuous DD, only if $%
Q\equiv \frac{t_f}{t_D}$ is an integer, then one might achieve the best
final fidelity. In the actual implementation, e.g., the (blue) solid line in
Fig.\thinspace 5, the peak-value of the final fidelity occurs for $Q\simeq
5.94$ instead of $Q=6$. This small deviation can be traced back to the
approximations made in the first-order Magnus expansion. Note also that the
associated DD field amplitude is $3.7$ in dimensionless units, whereas the
peak amplitude of the CD driving field is about $2.4$. By contrast, if we
boost the final fidelity to $F(t_{f})\approx 0.97$ by solely increasing the
sweeping rate in the CD driving without using DD, the peak strength of the
CD field will be as large as $15$. That is to say, the use of a DD field is
indeed helpful to battle against an environment in the realization of CD
driving.


\section{Concluding Remarks}

In this work we have investigated finite-time Landau-Zener processes in the
presence of an environment, which is modeled by a broadened cavity mode at
zero temperature. Such an environment model allows us to use the hierarchy
equation method to obtain numerically exact results of the dynamics without
any approximations.

The fidelity for the system to be found at the final target adiabatic state
shows a non-monotonic dependence on the system-bath coupling strength and on
the sweeping rate in the energy bias parameter. This signals an interesting
competition between the otherwise unitary LZ process and environment-induced
relaxation and excitation. More importantly, it is found that an environment
can affect the performance of CD driving significantly. If the LZ process is
not short enough, the addition of CD driving may lend more opportunities to
the environment to degrade the performance even further. It is also found
that different versions of CD driving can have different degrees of
robustness against environment effects.

Our results indicate that more studies of CD driving in the presence of an
environment will be fruitful. As a starting point, we discussed the
possibility of combining CD driving with dynamical decoupling (DD). At least
for the simple case we investigated, it is possible to combine CD and DD to
realize high-fidelity adiabatic manipulation in the presence of an
environment, without requiring the process to be extremely fast.

\section*{Acknowledgements}

Z.S. is supported by the National Natural Science Foundation of China under
Grants No. 11375003, the Program for HNUEYT under Grant No. 2011-01-011, the
Zhejiang Natural Science Foundation with Grant No. LZ13A040002, the funds
for the Hangzhou-City Quantum Information and Quantum Optics Innovation
Research Team. D.P. acknowledges support by the SUTD Start-Up Grant
EPD2012-045. D.P. and J.G. are also supported by Singapore MOE Academic
Research Fund Tier-2 project (Project No. MOE2014-T2-2-119, with WBS No.
R-144-000-350-112). J.G. also wishes to thank Y. Tanimura for teaching his
hierarchy equation method at NUS.


\begin{thebibliography}{99}
\bibitem{LZ} S. Shevchenko, S. Ashhab, F. Nori, Phys. Rep. \textbf{492}, 1
(2010).

\bibitem{Theoretical-1} S. Gasparinetti, P. Solinas, J. P. Pekola, Phys.
Rev. Lett. \textbf{107}, 207002 (2011).

\bibitem{Theoretical-2} C. Kasztelan, \textit{et al}., Phys. Rev. Lett.
\textbf{106}, 155302 (2011).

\bibitem{Theoretical-3} J.-N. Zhang, C. P. Sun, S. Yi, F. Nori, Phys. Rev. A
\textbf{83}, 033614 (2011).

\bibitem{Theoretical-4} A. Altland, V. Gurarie, Phys. Rev. Lett. \textbf{100}%
, 063602 (2008).

\bibitem{experiment2} Y.-A. Chen, S. D. Huber, S. Trotzky, I. Bloch, E.
Altman, Nature Phys. \textbf{7}, 61 (2011).

\bibitem{experiment3} M. G. Bason, \textit{et al}., Nature Phys. \textbf{8}
147 (2012).

\bibitem{atoms} D. M. Berns, \textit{et al}., Nature \textbf{455}, 51 (2008).

\bibitem{BEC} A. Zenesini \textit{et al}., Phys. Rev. Lett. \textbf{103},
090403 (2009).

\bibitem{super1} J.Q. You, F. Nori, Physics Today \textbf{58} (11), 42
(2005).

\bibitem{super2} J.Q. You, F. Nori, Nature \textbf{474}, 589 (2011).

\bibitem{super3} M. Sillanp\"{a}\"{a}, \textit{et al}., Phys. Rev. Lett.
\textbf{96}, 187002 (2006).

\bibitem{super4} G. Z. Sun, \textit{et al}., Nature Communications \textbf{1}
51 (2010).

\bibitem{Wubs} M. Wubs, \textit{et al}., New J. Phys. \textbf{7}, 218 (2005).

\bibitem{Keeling} J. Keeling, V. Gurarie, Phys. Rev. Lett. \textbf{101},
033001 (2008).

\bibitem{Zhe-LZ} Z. Sun, J. Ma, X. Wang, F. Nori, Phys. Rev. A \textbf{86},
012107 (2012).

\bibitem{Wubs-Saito-zeroT} M. Wubs, K. Saito, S. Kohler, P. H\"{a}nggi, Y.
Kayanuma, Phys. Rev. Lett. \textbf{97}, 200404 (2006).

\bibitem{Ind-crossing1} K. Saito, M. Wubs, S. Kohler, Y Kayanuma, P. H\"{a}%
nggi, Phys. Rev. B \textbf{75}, 214308 (2007).

\bibitem{LZ-dephasing} J. E. Avron, M. Fraas, G. M. Graf, P. Grech, Commun.
Math. Phys. \textbf{305}, 633--639 (2011).

\bibitem{LZ-Longwen} L. Zhou, D. Y. Tan, J. Gong, arXiv:1507.07331

\bibitem{Numerical-1} P. Nalbach, M. Thorwart, Phys. Rev. Lett. \textbf{103}%
, 220401 (2009).

\bibitem{Numerical-2} P. P. Orth, A. Imambekov, K. Le Hur, Phys. Rev. A
\textbf{82}, 032118 (2010).

\bibitem{Numerical-3} R. S. Whitney, M. Clusel, T. Ziman, Phys. Rev. Lett.
\textbf{107}, 210402 (2011).

\bibitem{Numerical-4} C. Xu, A. Poudel, M. G. Vavilov, Phys. Rev. A \textbf{%
89}, 052102 (2014).

\bibitem{Numerical-5} V. L. Pokrovsky, D. Sun, Phys. Rev. B \textbf{76},
024310 (2007).

\bibitem{Numerical-6} S. Javanbakht, P. Nalbach, M. Thorwart, Phys. Rev. A
\textbf{91}, 052103 (2015).

\bibitem{Numerical-7} A. Dodin, S. Garmon, L. Simine, D. Segal, J. Chem.
Phys. \textbf{140}, 124709 (2014).

\bibitem{Numerical-8} P. P. Orth, A. Imambekov, K. Le Hur, Phys. Rev. B
\textbf{87}, 014305 (2013).

\bibitem{Tanimura1991} Y.~Tanimura, P.~G.~Wolynes, Phys.~Rev.~A \textbf{43},
4131 (1991).

\bibitem{Tanimura2010JCP} M.~Tanaka, Y.~Tanimura, J.~Chem.~Phys. \textbf{132}
214502 (2010).

\bibitem{Tanimura2010PRL} A.~G.~Dijkstra, Y.~Tanimura, Phys.~Rev.~Lett.
\textbf{104}, 250401 (2010).

\bibitem{Ishizaki2007} A.~Ishizaki, Y.~Tanimura, J.~Phys.~Chem.~A \textbf{111%
}, 9269 (2007).

\bibitem{Tanimura2006} Y.~Tanimura, J.~Phys.~Soc.~Jpn. \textbf{75}, 082001
(2006).

\bibitem{YJ Yan-1} J. S. Jin, X. Zheng, Y. J. Yan, J.~Chem.~Phys. \textbf{122%
}, 041103 (2004).

\bibitem{YJ Yan-2} J.~Xu, R.-X.~Xu, Y.~J.~Yan, New J. Phys. \textbf{11},
105037 (2009).

\bibitem{Ishizaki} M.~Sarovar, A.~Ishizaki, G.~R.~Fleming, K.~B.~Whaley,
Nature Phys. \textbf{6}, 462 (2010);

\bibitem{MaJian-Hierarchy} J. Ma, Z. Sun, X. Wang, F. Nori, Phys. Rev. A 85,
062323 (2012).

\bibitem{Guo Wei-Berry phase} W. Guo, J. Ma, X. Yin, W. Zhong, X. Wang,
Phys. Rev. A \textbf{90}, 062133 (2014).

\bibitem{CD-1} E. Torrontegui, \textit{et al.}, Adv. At. Mol. Opt. Phy.,
\textbf{62} 117 (2013).

\bibitem{CD-Rice-1} M. Demirplak, S. A. Rice, J. Phys. Chem. A \textbf{107},
9937 (2003);

\bibitem{CD-Rice-2} M. Demirplak, S. A. Rice, J. Phys. Chem. B \textbf{109},
6838 (2005);

\bibitem{CD-Rice-3} M. Demirplak, S. A. Rice, J. Chem. Phys. \textbf{129},
154111 (2008).

\bibitem{CD-2 Berry} M. V. Berry, J. Phys. A: Math. Theory \textbf{42},
365303 (2009).

\bibitem{CD-4} S. Gu\'{e}rin, V. Hakobyan, H. R. Jauslin, Phys. Rev. A
\textbf{84}, 013423 (2011).

\bibitem{CD-5} S. Masuda, K. Nakamura, Phys. Rev. A \textbf{84}, 043434
(2011).

\bibitem{CD-6} X. Chen, I. Lizuain, A. Ruschhaupt, D. Gu\'{e}ry-Odelin, J.
G. Muga, Phys. Rev. Lett. \textbf{105}, 123003 (2010).

\bibitem{CD-7} S. Ib\'{a}\~{n}ez, X. Chen, E. Torrontegui, J.G. Muga, A.
Ruschhaupt, Phys. Rev. Lett. \textbf{109}, 100403 (2012).

\bibitem{CD-Harmonic-1} X. Chen, J. G. Muga, Phys. Rev. A \textbf{82},
053403 (2010).

\bibitem{CD-atom-trans} R. Bowler, \textit{et al.}, Phys. Rev. Lett. \textbf{%
109}, 080502 (2012).

\bibitem{CD-quant-comp} M. S. Sarandy, E. I. Duzzioni, R. M. Serra, Phys.
Lett. A \textbf{375}, 3343 (2011).

\bibitem{CD-quant-simu} H.-K. Lau, D. F. V. James, Phys. Rev. A \textbf{85},
062329 (2012).

\bibitem{CD-manybody} A. del Campo, Phys. Rev. Lett. \textbf{111}, 100502
(2013).

\bibitem{CD-cooling} Y. Li, L. A. Wu, Z. D. Wang, Phys. Rev. A \textbf{83},
043804 (2011).

\bibitem{DelCampoP2014} A. del Campo, J. Goold, M. Paternostro,
Sci. Rep. 4, 6208 (2014).

\bibitem{ZhengPoletti2015} Y. Zheng S. Campbell, G. De Chiara, D. Poletti,
arxiv:1509.01882 (2015).

\bibitem{CD-open-1} G. Vacanti, R. Fazio, S. Montangero, G. M. Palma, M.
Paternostro, V. Vedral, New J. Phys. \textbf{16}, 053017 (2014).



\bibitem{CD-open-4} Y. H. Issoufa, A. Messikh, Phys. Rev. A \textbf{90},
055402 (2014).


\bibitem{viola1} L. Viola, S. Lloyd, Phys. Rev. A \textbf{58}, 2733 (1998).

\bibitem{viola2} L. Viola, E. Knill, S. Lloyd, Phys. Rev. Lett. \textbf{82},
2417 (1999).

\bibitem{Lorentz-quant-zero} S. Maniscalco, \textit{et al., }Phys. Rev.
Lett. \textbf{100}, 090503 (2008).

\bibitem{DD-2} F. F. Fanchini, J. E. M. Hornos, and R. d. J. Napolitano,
Phys. Rev. A \textbf{75}, 022329 (2007).

\bibitem{DD-3} F. F. Fanchini, J. E. M. Hornos, and R. d. J. Napolitano,
Phys. Rev. A \textbf{76}, 032319 (2007).

\bibitem{DD-4} X. Xu, \textit{et al}., Phys. Rev. Lett. \textbf{109}, 070502
(2012).

\bibitem{DD-Adam-Gong} A. Z. Chaudhry and J. Gong, Phys. Rev. A \textbf{85},
012315 (2012).

\bibitem{Magnus-expansion} W. Magnus, Commun. Pure Appl. Math. \textbf{7},
649 (1954).
\end{thebibliography}
\end{document}